# Tidal Evolution of Close-in Extra-Solar Planets

by

Brian Jackson, Richard Greenberg, and Rory Barnes

Lunar and Planetary Laboratory
University of Arizona
1629 E University Blvd, Tucson AZ 85721-0092

Running Title: Tidal Evolution

Abstract

The distribution of eccentricities $e$ of extra-solar planets with semi-major axes $a$ > 0.2 AU is very uniform, and values for $e$ are relatively large, averaging 0.3 and broadly distributed up to near 1. For $a$ < 0.2 AU, eccentricities are much smaller (most $e$ < 0.2), a characteristic widely attributed to damping by tides after the planets formed and the protoplanetary gas disk dissipated. Most previous estimates of the tidal damping considered the tides raised on the planets, but ignored the tides raised on the stars. Most also assumed specific values for the planets' poorly constrained tidal dissipation parameter $Q_p$. Perhaps most important, in many studies, the strongly coupled evolution between $e$ and $a$ was ignored. We have now integrated the coupled tidal evolution equations for $e$ and $a$ over the estimated age of each planet, and confirmed that the distribution of initial $e$ values of close-in planets matches that of the general population for reasonable $Q$ values, with the best fits for stellar and planetary $Q$ being $\sim 10^{5.5}$ and $\sim 10^{6.5}$, respectively. The accompanying evolution of $a$ values shows most close-in planets had significantly larger $a$ at the start of tidal migration. The earlier gas disk migration did not bring all planets to their current orbits. The current small values of $a$ were only reached gradually due to tides over the lifetimes of the planets. These results may have important implications for planet formation models, atmospheric models of "hot Jupiters", and the success of transit surveys.

## 1. Introduction

Over 200 planets have been discovered outside of our solar system (*e.g.,* Marcy *et al.* 2005; Butler *et al.* 2006). Each new discovery contributes to the rapidly evolving knowledge about planetary formation and evolution processes. However, current observational methods favor the detection of massive extra-solar planets close to their host star. Thus most of the extra-solar planets discovered so far have a mass comparable to, or exceeding, Jupiter's mass and are presumably gas giants composed largely of hydrogen, and most have orbital semi-major axes less than 10 AU, with about 20% having a semi-major axis less than 0.2 AU.

Figure 1 shows the semi-major axes $a$ and eccentricities $e$ of all observed extra-solar planets. Eccentricities of extra-solar planets with $a > 0.2$ AU are relatively large, averaging 0.3 and broadly distributed up to near 1. The large values seem to run contrary to the prevailing belief that planets form embedded in a protoplanetary gas disk, which would tend to damp out eccentricities through gas and particle drag (*e.g.*, Lissauer 1993; Ida & Lin 2004). The distribution of eccentricities for planets is fairly uniform over $a$, for $a > 0.2$ AU. A Kolmolgorov-Smirnov (K-S) test (Press *et al.* 1996) shows that the $e$ distribution for $a$ between 0.2 and 1.0 AU matches that for $a$ between 1.0 and 5.0 AU at the 96% confidence level. For close-in extra-solar planets (by which we mean $a \leq$ 0.2 AU), $e$ tends to be less, although even there the average $e$, 0.09, is larger than is typical for our solar system. For this very different $e$ distribution, the K-S test shows agreement at only the 0.1% level compared with planets further out.

Rasio *et al.* (1996) suggested that tides raised between close-in planets and their host stars might help explain their relatively low $e$ values. Because the magnitude of tidal effects falls off very rapidly with increasing $a$, the action of tides could plausibly have reduced $e$ for close-in planets and not for those farther out. Even with that explanation, there remains the question of why extra-solar planetary systems formed with such large eccentricities. It is possible that whatever process generated such large eccentricities was less effective for close-in extra-solar planets. However, here we consider the conventional idea that close-in planets began with a distribution of *e* similar to that of planets farther out. We use that distribution to constrain tidal parameters and evolution rates.

Bear in mind, however, that the *e* value for any extra-solar planet can vary by orders of magnitude over secular timescales, due to the dynamical interactions between multiple planets in the system (Barnes & Greenberg 2006). Therefore any conclusions drawn based on the current *e* distribution are subject to uncertainties associated with these interactions. We assume that the statistical sample is adequate enough that these variations average out.

Because the rates of change of *e* and *a* depend on the physical properties of the planet and its host star, we can compare computations of tidal evolution (based upon various assumptions) with the observed *e* distribution (for $a > 0.2$ AU), using estimates of a system's age, to constrain those properties. Previous work has employed tidal considerations in attempts to constrain extra-solar planetary masses $M_p$ and radii $R_p$ and has investigated tidal heating rates.

Trilling (2000) considered what limits might be placed on a close-in planet's mass if its *e* is observed to be greater than zero. That study used the fact that the smaller the value of $M_p$, the more quickly the tide raised on a planet by its star can circularize the orbit. Lower limits were derived on the masses of some planets based on the observation that they still have substantial eccentricities. Bodenheimer *et al.* (2003) applied a similar argument to constrain planetary radii. The rate of circularization due to tides raised on a planet increases with a planet's radius, so in principle upper limits could be found for the radius if *e* has not damped down. Bodenheimer *et al.* (2003) also modeled the effect of tidal heating for HD 209458 b to try to explain the surprisingly large observed radius of 1.27 Jupiter radii (Charbonneau *et al.* 2000). They found that, unless HD 209458 b has no core, tidal heating at the current *e* is insufficient to explain the large radius.

The constraints provided by those two investigations are questionable for several reasons. First, they depend on assumed values of the tidal dissipation parameter $Q_p$ for the planet. Estimates of $Q_p$ are usually taken from highly uncertain estimates based on models of the tidal evolution the Galilean satellite system (Yoder & Peale 1982; Greenberg 1982, 1989) or on models of dissipation within the planet (Goldreich & Nicholson 1977; Ogilvie & Lin 2004). Second, they neglected the orbit-circularizing effects of tides raised on the host star by the close-in extra-solar planet. Third, they ignored the strong coupling of the tidal evolution of $e$ with that of $a$. In our study, rather than assuming values for stellar and planetary $Q$, we use tidal evolution rates, along with conservative assumptions about $M_p$ and $R_p$, to test the tidal circularization hypothesis and, in the process, constrain $Q_p$ and $Q_*$.

To test the hypothesis that tides have been responsible for reducing $e$, we numerically integrate the canonical tidal evolution equations (Goldreich & Soter 1966) backwards in time for all planets closer than 0.2 AU for which we have adequate information. This range of $a$ values includes all planets for which tidal evolution could conceivably be significant. For each planet, we begin the integrations at the current best estimates of eccentricity and semi-major axis, $e_{current}$ and $a_{current}$, and integrate backwards to the orbital elements, $e_{initial}$ and $a_{initial}$, at the time tides began to dominate the orbital evolution. We assume tidal evolution began when the protoplanetary disk had dissipated and collisional effects became negligible.

We include the effects of the tide raised on the star as well as on the planet since the effects of both can be important (as shown below). We assume that a typical value of $Q_p$ applies to all planets and a typical value of $Q_*$ applies to all stars. *A priori* estimates of these parameters are uncertain, so we repeated the full set of calculations for various pairs of values of $Q_p$ and $Q_*$.

Specific stars and planets may have different values, but we assume that the evolution of the population as a whole is not affected by such individual variations.

For each pair of $Q_p$ and $Q_*$, our integrations yield a distribution of values of $e_{initial}$. We then compare the *computed* distribution of $e_{initial}$ for the close-in planets to the *observed* distribution of $e$ for farther-out extra-solar planets. We determine which pair of tidal dissipation parameters gives the best fit of the computed distribution to the observed one. As we shall show, the $Q$ values obtained in this way are quite reasonable and consistent with previous evidence. The fact that such plausible parameters lead to a match between the computed initial $e$ distributions for close-in planets and the observed $e$ distribution for farther-out planets seems to confirm the tidal circularization hypothesis. Moreover, our results demonstrate that, for several close-in extra-solar planets, stellar and planetary tides have significantly reduced semi-major axes (as well as eccentricities) *after* the planets formed and gas disk migration ceased (Terquem, Papaloizou & Lin 1999).

In Section 2, we review the physics and dynamical equations governing tidal evolution, and we discuss previous studies of tidal evolution. In Section 3, we discuss the details of our methodology and the assumptions involved. In Section 4, we present the results of our tidal modeling, and in Section 5, we discuss their implications and significance. In Section 6, we summarize our results and discuss caveats of our study.

## 2. Tidal Evolution

### 2.1. Tidal Theory

Planetary tidal evolution was modeled mathematically by Darwin (1908), Jeffreys (1961), Goldreich (1963), Macdonald (1964) and many others. Useful equations, were compiled by Goldreich & Soter (1966) and Kaula (1968). Recasting them in a form appropriate for close-in extra-solar planets yields:

$$\frac{1}{e}\frac{de}{dt} = -\left(\frac{63}{4}(GM_*^{\ 3})^{1/2}\frac{R_p^{\ 5}}{Q_p M_p} + \frac{171}{16}(G/M_*)^{1/2}\frac{R_*^{\ 5}M_p}{Q_*}\right)a^{-13/2} \tag{1}$$

$$\frac{1}{a}\frac{da}{dt} = -\left(\frac{63}{2}(GM_*^{\ 3})^{1/2}\frac{R_p^{\ 5}}{Q_p M_p}e^2 + \frac{9}{2}(G/M_*)^{1/2}\frac{R_*^{\ 5}M_p}{Q_*}\right)a^{-13/2}\ . \tag{2}$$

Here $G$ is the gravitational constant, $R$ is a body's radius, $M$ its mass and $Q$ its tidal dissipation parameter, and subscripts $p$ and $*$ refer to the planet and star, respectively.

The coefficients in these equations include a factor representing the tidal Love number $k$ of each distorted body. As written above, the numerical coefficients are what they would be if $k$ = 3/2. However, the actual values of the Love numbers are unknown, depending on the tidal-effective rigidity of the body, the radial density distribution, etc. Hence, we incorporate into our definition of $Q$ a correction factor for $k$. Thus what we call $Q$ is similar to the $Q'$ of Goldreich & Soter (1966). These equations describe the orbit-averaged effects of the tides. The effects of the tide raised on the star by the planet are reflected in the terms involving $Q_*$ (which we call the stellar tide), while the terms involving $Q_p$ reflect the effect of the tide raised on the planet by the star (which we call the planetary tide).

Complex dissipative processes in each body result in a phase lag between the tidal forcing potential and the body's deformation. In analogy to a damped, driven harmonic oscillator, the phase lag angle is given by $1/Q$, in the limit of large $Q$. The equations above assume that all Fourier components of the tide have an equal phase lag, that is $Q$ is independent of frequency. Alternative assumptions are possible (*e.g.,* Hut 1981; Eggleton *et al.* 1998), and the exact nature of a body's tidal response remains under investigation (Hubbard 1974; Goldreich & Nicholson 1977; Ogilvie &

Lin 2004; Ogilvie & Lin 2007). Unless and until a new consensus emerges, Equations (1) and (2) represent a reasonable and widely adopted representation of tidal evolution.

Equations (1) and (2) assume that *e* values are small. Since, for close-in extra-solar planets, *e* was much larger in the past, higher-order corrections may be important. Several authors have worked out tidal models for arbitrary *e* (Zahn 1977; Hut 1981; Eggleton *et al.* 1998; Mardling & Lin 2002). However, their results require specific assumptions about the response of a body to tidal forcing, the nature of which is uncertain. For example, in the case of larger eccentricities, harmonics of various frequencies come into play (*e.g.* Goldreich 1963), so formulations will depend more strongly on what underlying and implicit frequency dependence is assumed for the tidal response. In any case, these methods predict faster tidal evolution for large *e*, so our model should provide conservative estimates of the rate of tidal evolution. Future work should investigate the effects of alternative assumptions on tidal circularization and will consider higher order corrections to the tidal equations, but our approach is a reasonable first-cut at testing the tidal circularization hypothesis for the population of close-in extra-solar planets, with coupled equations for changes in *a* and *e*.

Equations (1) and (2) also assume that the planet's orbital period is short compared with the star's rotation period. The second term in Equation (1) requires that the host star's rotation period $P_*$ is greater than 2/3 the orbital period $P_p$ of the tidally evolving planet, while the second term in Equation (2) assumes $P_* > P_p$. If either condition were violated, the corresponding term would change sign (Goldreich & Soter 1966). Dobbs-Dixon *et al.* (2004) computed the tidal evolution of planets considering the possibility that their orbital periods might be close to their star's rotational period.  Their analysis required a model that incorporated changes in effective *Q* with frequency, because primary tidal components could undergo drastic changes in frequency. Here we assume that the rotation of each star is slow enough compared with the orbital motion that Equations (1) and (2) remain valid.

Is the assumption of sufficiently large $P_*$ reasonable? The rotation of young, rapidly rotating stars is slowed initially due to loss of angular momentum to the circumstellar disk through magnetic coupling between the star and the disk (Tinker *et al.* 2002). This effect can slow the young's stars rotation to a period $\geq$ 10 days in a few Myr. After dissipation of the circumstellar disk, the stellar rotation continues to slow due to shedding of angular momentum through the stellar wind (Skumanich 1972; Verbunt & Zwaan 1981; Ogilvie & Lin 2007), which also helps to keep the star's rotation longer than the revolution period of most close-in planets. Trilling (2000) and Barnes (2001) list some stellar rotation periods, corroborating our assumption, with two possible exceptions. In our models, the longest-period planet that experiences any significant tidal evolution is HD 38529 b, with a period of 14 days. The rotation period for its star is unknown, but many extra-solar host stars have rotation periods > 14 days, so we used the signs in Equations (1) and (2). The other exception is $\tau$ Boo b, with $P_*$ = 3.2 (Henry *et al.* 2000) and $P_p$ = 3.31 days (Butler *et al.* 2006). However, in light of the uncertainties in the measurement of $P_*$, we include $\tau$ Boo b here anyway.

Equations (1) and (2) also assume the planet is rotating nearly synchronously with its orbit. This assumption is reasonable because any close-in planet should have spun down to near synchronous rotation in ~ 1 Myr (Peale 1977; Rasio *et al.* 1996), too early to affect tidal orbital evolution, which takes place over billions of years. (There is a small probability that some planets can become trapped in a non-synchronous spin-orbit resonance (Winn & Holman 2005), but for our purposes we assume the probability is negligibly small.) With synchronous rotation, there is no net exchange of angular momentum between the planet's rotation and its orbit. The $Q_p$ terms in Equations (1) and (2) conserve angular momentum (to second order in *e*).

For many short-period extra-solar planets, tides raised on the planet by the star dominate the tidal evolution of the planet (*i.e.* in Equations [1] and [2], $Q_p$ terms dominate). However, given uncertainties in the values of various parameters, the effects of the tide raised on the star cannot be neglected *a priori*, and indeed our results show they can be very important. Equations (1) and (2) show the strong non-linear coupling between *e* and *a*. As a general rule, the equations cannot legitimately be treated separately. When properly considered together, Equations (1) and (2) cannot be solved analytically for *e* and *a* as functions of time. Adams & Laughlin (2006) considered approximate solutions that may apply in some circumstances. We have however solved for *e* as a function of *a*:

$$(E_p + E_*)\ln(a/a_0) = \tfrac{1}{2}A_p(e^2 - e_0^{\,2}) + A_*\ln(e/e_0) \qquad (3)$$

where $E_p = (63/4)\,(GM_*^3)^{1/2}\,R_p^5/(Q_p\,M_p)$, $A_p = (63/2)\,(GM_*^3)^{1/2}\,R_p^5/(Q_p\,M_p)$, $E_* = (171/16)\,(G/M_*)^{1/2}\,R_*^5\,M_p/Q_*$ and $A_* = (9/2)\,(G/M_*)^{1/2}\,R_*^5\,M_p/Q_*$. The constants $e_0$ and $a_0$ refer to some values at some time $t = 0$. However, to find the changes in *e* and *a* with time, numerical integration is required, and that is our method.

## 2.2. Previous Studies of Tides Related to Extra-Solar Planets

Most previous work has employed only selected terms from Equations (1) and (2) in the effort to explain the low eccentricities among close-in extra-solar planets. Rasio *et al.* (1996), Trilling (2000), and Bodenheimer *et al.* (2003) all took *a* to be constant in Equation (1), neglecting the variation given by Equation (2), and neglected the circularizing effect of the stellar tide (the second term in Equation [1]). With these approximations, Equation (1) becomes

$$\frac{1}{e}\frac{de}{dt} = -\left(\frac{63}{4}\left(GM_*^{\ 3}\right)^{1/2}\frac{R_p^{\ 5}}{Q_P M_p}\right)a^{-13/2} \equiv -\, 1/\tau_{circ} \ . \qquad (4)$$

The solution of Equation (4) is an exponential damping of $e$ on the timescale $\tau_{circ}$.

Trilling (2000) proposed that, if an extra-solar planet's orbit currently has a significant eccentricity (Trilling (2000) chose $e > 0.1$), then $\tau_{circ}$ must be longer than the age of the system ($\tau_{age}$), because otherwise $e$ would have damped down by now. Trilling (2000) assumed $R_p$ = 1.27 Jupiter radii, independent of $M_p$, which is reasonable for gas giants with $0.3\ M_J < M_p < 10\ M_J$ (where $M_J$ is Jupiter's mass), because the polytropic equation of state for hydrogen implies the internal density is proportional to the planet's mass (Hubbard 1984). Trilling (2000) also adopted the value $Q_p = 10^5$ and the best available determinations of $M_*$, $a$, and $e$. Then, using Equation (4), Trilling (2000) calculated the minimum $M_p$ for several planets, based on the assumption that $\tau_{circ} > \tau_{age}$.

Figure 2 illustrates that approach for the extra-solar planet HD 217107 b ($e$ is currently 0.14), showing the relationship between $\tau_{circ}$ and $M_p$. The estimated age of HD 217107 b is ~ 6 Gyr (Takeda *et al.* 2007), which led Trilling (2000) to infer that $M_p$ must be greater than 4 $M_J$ in order to satisfy $\tau_{circ} > \tau_{age}$ (Figure 2). However, the picture changes significantly if we include the effect of the tide raised on the star, setting $Q_* = 10^5$ and $R_*$ = 1.1 solar radii (Valenti & Fischer 2005), but still retaining the assumption of constant $a$. Since the effect of the stellar tide increases as $M_p$ increases, the circularization timescale now drops as $M_p$ grows larger (Figure 2). In this case, Figure 2 shows that $\tau_{circ}$ is smaller than $\tau_{age}$ for any value of the planet's mass. Thus, if we were to make inferences simply on the basis of a comparison of $\tau_{circ}$ with $\tau_{age}$, the implication would be that $e$ should have damped to zero long ago.

Bodenheimer *et al.* (2003) applied logic similar to Trilling's in order to constrain $R_p$, assuming the radial-velocity estimated minimum mass for $M_p$. Those results suffer from the same approximation used in Trilling (2000): the neglect of tides raised on the star. And both studies assumed $a$ is constant on the right side of Equation (1). In addition, both studies assumed a value *a priori* for the poorly constrained $Q_p$ (Trilling (2000) assumed $Q_p \sim 10^5$, Bodenheimer *et al.*(2003) $Q_p \sim 10^6$), which introduced further uncertainty.

Previous considerations of $Q$ values have shown that constraints are fairly weak, and the mechanisms of tidal friction remain only partially understood, although some progress has been made. For constraints on planetary $Q_p$, it is reasonable to refer, as Bodenheimer *et al*. (2003) and Trilling (2000) did, to studies of Jupiter. Yoder & Peale (1981) proposed that $Q$ for Jupiter lay in the range $6\times10^4$ to $2\times10^6$. The range comes from consideration of the role of tides on Jupiter in the evolution of its satellite Io. If Jupiter's $Q$ were smaller than this range (remember small $Q$ means strong dissipation), Io could not be as close to Jupiter as it is, even taking into account the exchange of orbital energy and angular momentum with the other satellites in the Laplace resonance. However, the upper limit is much more model dependent. It assumes that the Laplace resonance is in a steady-state that involves a balance between effects of tides raised on Io by Jupiter and those raised on Jupiter by Io. We have a reasonable idea of the very small value of $Q$ for Io, based on its observed volcanism and strong thermal emission. Thus, dissipation must be high enough in Jupiter ($Q < 2\times10^6$) to maintain the steady-state assumed by Yoder & Peale (1981).

However, it is equally plausible that the Jupiter system is not in that steady-state (Greenberg 1982, 1989), but that tides on Io are currently dominant. In fact, a model of the formation of the Jovian system by Peale & Lee (2002) also suggests that the resonance has evolved over the long term in the way expected if tides on Io dominate. Perhaps more significantly, observations of the orbital evolution of Io also show a rate consistent with dominance of tides on Io and negligible tidal dissipation in Jupiter (Aksnes & Franklin 2001).

More recently, Ogilvie & Lin (2004) modeled the fluid dynamical problem of a tidally driven gaseous planet. Their results suggested a complex dependence of $Q_p$ on the tidal forcing frequency and the planet's internal stratification. They conclude that a constant $Q_p$ does not adequately capture the full tidal dissipation behavior of gas giant planets, but that the effective $Q_p$ values for close-in extra-solar planets are of order $5\times10^6$.

Turning next to stars, we find that constraints on $Q_*$ have been derived from observations of tidally evolved binary stars and from theoretical modeling of stellar tidal dissipation. Mathieu (1994) observed that many main sequence binary stars with orbital periods of 10 days or less have small or zero *e*. Lin *et al.* (1996) drew on this observation and inferred a value for $Q_* \sim 10^5$. Carone & Patzöld (2007) derived a value $3\times10^7 < Q_* < 2.25\times10^9$ for OGLE-TR-56 b. Ogilvie & Lin (2007) modeled stellar tidal dissipation numerically and found (as they did for tides on a planet) a complex dependence of $Q_*$ on tidal forcing frequency and on the mode of dissipation. Taken altogether, these studies illustrate the uncertainty inherent in assuming any values of $Q_*$, as well as $Q_p$.

It would be inappropriate, except in special cases, to ignore the effect of the stellar tide, which is equivalent to letting $1/Q_* = 0$. Some previous studies (*e.g.*, Ford & Rasio 2006) assumed that, as a general rule, during tidal evolution of extra-solar planets, orbital angular momentum is conserved. However, that is only true if the tide on the star is negligible. Tides raised on a star transfer angular momentum between the star's rotation and the planet's orbit. Thus interpretations of the orbital distribution among close-in planets that involve conservation of orbital angular momentum during tidal evolution should be regarded with caution.

Next we consider the common assumption that tidal variation of e can be estimated by holding *a* constant, which leads to Equation (4). To illustrate the effect of holding *a* constant, we numerically integrated Equations (1) and (2) backward in time for the planet τ Boo b for various *Q* values. For comparison, we show the corresponding exponential solutions of Equations (1) with *a* held constant. The results shown in Figure 3 reveal that ignoring changes in *a* does not give a good approximation to the actual solution.

In Figure 3, for $Q_* = 10^7$, we see that the behavior of *e* based on Equations (1) and (2) (solid curve) follows closely the exponential solution of Equation (1) with constant *a* (dashed line) for about the first Gyr back in time. This close agreement is reasonable, because the actual change in *a* is small during this time. But further back in time, we see that *a* begins to change significantly. Consequently, the assumption of constant *a* is no longer accurate, and the behavior of *e* (solid curve) begins to deviate from the exponential solution (dashed line). For smaller values of $Q_*$ (that is more dissipation), *a* varies even more quickly, so the approximation of constant *a* is inaccurate. Consequently the time variation of *e* diverges from the exponential solution. The commonly made assumption that variation of *a* can be ignored in Equation (1) is not accurate and can lead to incorrect conclusions.

The complex behavior of *e* (*i.e.* complex relative to the exponential solution) leads to a surprising relationship between $Q_*$ and the change in the value of *e* over time. In Figure 3 we see that the total change in *e* over 15 Gyr decreases as $Q_*$ increases from $10^4$ to $10^6$. That result seems reasonable because tidal effects generally decrease with increasing $Q_*$. However, for the larger $Q_*$ value (*e.g.*, $10^7$), the change in $e$ is much greater than any of the cases with smaller $Q_*$. The reason is that $a$ changes less, spending more time at low values, which means tidal effects are stronger. Clearly the behavior of $e$ is very different from what would have been predicted by assuming $a$ to be constant in Equation (1).

We plot $e_{intial}$ for τ Boo b as a function of both $Q_p$ and $Q_*$ in Figure 4. Once $Q_*$ is greater than about $10^7$, the evolution is independent of tides raised on the star, and for $Q_p > 10^6$, it is independent of tides raised on the planet. In general, increasing either $Q$ (decreasing tidal dissipation) tends to decrease the change in $e$, as one might expect. However, the "heel" of the curves (around $Q_p \sim 10^4$ and $Q_* \sim 10^3$) shows the reversal of this trend that we saw in Figure 3. For example, if $Q_p$ is between $10^3$ and $10^6$, as $Q_*$ increases from $10^2$, the change in $e$ first decreases (as expected intuitively), but then increases. As shown in Figure 3, this phenomenon is explained by the concurrent change in $a$.

This discussion demonstrates the importance of retaining all the terms in Equations (1) and (2) and of considering the coupled evolution of $a$ and $e$ given by those equations. Specific terms can only be ignored under special circumstances. Incorporating the effect of the tide on the star and the tidal variation of $a$ significantly improves the accuracy of tidal calculations and, in many cases, leads to qualitatively different behavior.

## 3. Method

We numerically integrated Equations (1) and (2) for all planets with semi-major axes less than 0.2 AU for which we have adequate data (Table 1). Tidal evolution is negligible (less than a few percent change in $e$ or $a$) for any plausible $Q$ values if $a$ is greater than 0.2 AU. For each planet we ran the integration backward in time from the present to 15 Gyr ago. Also for each planet, we repeated the integration for 289 combinations of $Q_p$ and $Q_*$, evenly distributed over 17 values of each $Q$ from $10^4$ to $10^8$ in increments of $10^{0.25}$.

Our sample of planets, listed in Table 1, excludes those for which $e$ is unknown, although often such planets have $e$ tabulated as zero. We also restrict our study to planets for which there is some estimate available for the age of the system. Otherwise we have no way to determine the time of the initial $e$ and $a$ values.  Age estimates for solar-type stars are difficult and often uncertain (see, *e.g.*, Saffe *et al.* 2006), but we use the estimates currently available (Table 1). For many systems, only a minimum or maximum age is available. If we could find only a minimum or maximum age, we used that value; if we found both, we took the average.

In choosing the age of the star as the age of the planetary system, we are assuming that tides began to dominate orbital evolution of close-in planets shortly after their host stars formed. Gas disk migration is thought to bring newly-forming planets inward for ~ 1 Myr (Chambers 2007), until the gas disk dissipates (Hillenbrand *et al.* 1998). Thus that migration process is completed on a timescale that is very short compared with tidal evolution becomes important, and we may assume that age of the system is equal to the age of the star.

The values for $M_p$ and $R_p$ come from various sources (see Table 1). Where a minimum mass ($M \sin i$) is available from radial-velocity measurements, we set $M_p$ equal to that value. This assumption should only contribute a small error (~30%, typically) due to uncertainty regarding the orbital inclination $i$ relative to the observer. Three of the planets we considered have been observed by stellar transit: GJ 436 b (Deming *et al.* 2007), HD 209458 b (Laughlin *et al.* 2005a), and HAT-2-P b (Bakos *et al.* 2007), so for these planets, we have used the directly measured values of $R_p$ and $M_p$. For other planets, our adopted $R_p$ depends on $M_p$. For $M_p \geq 0.3\ M_J$, we fix $R_p$ = 1.2 Jupiter radii, independent of mass. This radius represents the average radius of almost all of the close-in planets which have been observed by stellar transit (including those with nominal $e$ = 0, which we have not included in our tidal-evolution study), and this value is within 10% of nearly all their radii. (We exclude HD 149026 b from this average because its internal structure may be anomalous [Burrows *et al.* 2007]). Based on internal modeling, it is reasonable for $R_p$ to be independent of $M_p$ over this mass range. For planets with a minimum mass less than 0.3 Jupiter masses, we assume the planet has the same density as Jupiter and scale the radius accordingly. This assumption roughly agrees with the known $R_p$ for Uranus, Neptune, and GJ 436 b (Gillon *et al.* 2007; Deming *et al.* 2007).

Planetary radius is a sensitive function of many factors, including internal structure, thermal history, and atmospheric opacity. Moreover, radiative cooling would tend to reduce $R_p$ (Burrows *et al.* 2007), while tidal heating may counteract that effect to some degree. Our assumptions for $R_p$ can be revised as improved information becomes available but are sufficiently accurate for the current study.

Values for $R_*$ and $M_*$ also come from various sources, as referenced in Table 1. In five cases, radii are not given explicitly, but we have computed them from published values of surface gravity. For two stars, we computed $R_*$ using the empirical relation between $M_*$ and $R_*$ reported by Gorda & Svechnikov (1996).

4. Results

The 289 combinations of $Q_p$ and $Q_*$ that we tested (evenly spanning the range of $10^4$ to $10^8$ for each $Q$) gave a wide variety of distributions of orbital eccentricity for the close-in planets at the time that their tidal evolution began. Figure 5 shows these computed "initial" distributions of orbital elements, as well as the current distribution, for four examples of pairs of $Q$ values. One example, with $Q_p$ =$10^6$ and $Q_*$ = $10^5$, (bottom right) represents $Q$ values that have often been adopted in previous studies of extra-solar tidal evolution (Section 2.1). The other three examples represent some of the better fits of the initial *e* distribution of close-in planets to the distribution farther out.

In Figure 5, the white squares represent the current orbital elements of close-in planets, while the black triangles show the "initial" orbital elements based on the solution to the tidal-evolution equations. In cases where a black triangle is co-located with a white square, the equations of tidal evolution simply gave negligible changes in *a* or *e*. Solid black squares represent orbits outside 0.2 AU where we were confident that no tidal evolution could occur. Where the equations have artificially driven *e* backwards to values > 1, the meaningful interpretation is that *e* would be very large; we plot those values as $e = 1$.

Even a qualitative inspection of these examples demonstrates the varying degree of agreement with the *e* distribution farther out. The commonly assumed set of $Q_p$ and $Q_*$ exponents (6 and 5, respectively) does not appear to be a good fit. In that case, among the backwards-evolved close-in planets (black triangles), there is clearly a great excess of large initial *e* values (e.g., > 0.7) compared with planets farther out (the black squares). For the *Q* exponents 6.5 and 4.25, there is a distinct paucity of initial *e* values between 0.5 and 1, and a cluster at 1. This case, too, does not fit the *e* distribution for planets with larger *a* values. For the *Q* exponents 6.5 and 6.75, we see among the black triangles an excess of initial *e* values between 0.5 and 0.7, and too few with $e < 0.4$. Most of the 289 combinations of *Q* values give far worse fits than any of these four examples. Of these four examples, the combination *Q* exponents 6.5 and 5.5 appears to give the best qualitative fit to the population outside $a = 0.2$ AU.

A quantitative measure of the degree of agreement between, on one hand, the distribution of the computed initial values of *e* for the close-in planets and, on the other hand, the *e*-distribution observed for farther-out (greater *a*) planets confirms this qualitative impression. Recall from Section 1 that the K-S test showed the *e*-distribution to be quite consistent over a broad range of *a* values; the distribution of *e* for orbits with *a* from 0.2 to 1 AU is consistent with those from 1 to 5 AU at the 96% level of confidence. So now we can compare our computed initial *e* distribution for the close-in planets with the standard *e* distribution observed for $a > 0.2$ AU. Figure 6 shows the K-S scores as a function of the two *Q* values. Here we see reasonably good fits are possible only if $Q_p \approx 10^{6.5}$. Peaks in the K-S score are found with values ~70% for $Q_* = 10^{4.25}$ or $>10^{6.75}$. The best fit however, with a score of nearly 90%, is obtained for the case with $Q_p = 10^{6.5}$ and $Q_* = 10^{5.5}$.

Figure 7 shows the tidal evolution of *a* and *e* over time for each of the planets, in the case of the best-fit *Q* values. The current observed values are at the lower left end of each trajectory in (*a*, *e*) space. These points correspond to white squares in Figure 5. The tick marks show the position at intervals of 500 Myr, going back in time from the present toward the upper right. Black dots have been placed at a point representing the best age estimate for the planetary system. These points are shown as the triangles in Figure 5 (top left panel). We have indicated the evolution going back further in time for use if a more reliable age becomes available, or to indicate the evolutionary path if the *Q* values were different (but still with the same ratio assumed here).

The curvature of these trajectories is indicative of the relative importance of the terms in Equations (1) and (2). A trajectory that is concave to the upper left usually indicates that tides raised on the star are most important, while concavity to the lower right indicates that tides on the planet are the most important.

As *e* grows, a trajectory initially dominated by the stellar tide can become dominated by the planetary tide. The evolutionary track of 51 Peg b (currently at $a = 0.0527$ AU and $e = 0.013$) illustrates this behavior, reversing concavity around $e = 0.35$ and $a = 0.06$ AU. The dominance reversal is due to the first term in Equation (1), which depends on $e^2$: for large $e$, this term dominates $da/dt$. This effect underlines the importance of incorporating all the terms in Equations (1) and (2) when modeling tidal evolution.

Another important effect evident in Figure 5 is the acceleration or deceleration of tidal evolution as $e$ and $a$ decreases. The rate of tidal evolution is reflected in the spacing of tick marks along each trajectory. Moving forward in time from large values of $e$ and $a$ in the past, tidal evolution for many planets accelerates and then decelerates. For example, for 51 Peg b, the tick marks are closely spaced near the top of the trajectory, widely spaced in the middle, and closely spaced again near the bottom (near the present time). To understand this effect, consider tidal evolution moving forward in time. For larger values of $a$ (far in the past), tidal evolution proceeds slowly, since both $de/dt$ and $da/dt \propto a^{-13/2}$. However, as $a$ decreases, the rate of tidal evolution increases. Later, as $e$ becomes small enough, the rate of evolution again decreases.Although, currently, tidal evolution for many planets is slower than in the past, some bodies are still undergoing very rapid tidal evolution, such as HD 41004B b. However, with $M_p = 18\ M_J$, HD 41004B b is probably not a planet at all, but rather a brown dwarf. In this case, its history and evolution may be different from planets and may also have been affected by perturbations from other bodies in the system (Zhang & Hamilton 2007).

The evolutionary histories derived here include substantial changes in semi-major axis coupled with the changes in eccentricity. Figure 8 compares the initial value of $a$ with the current value. The initial value is the value of $a$ just after other orbit changing effects, such as gas drag or other effects of the planet-formation process, became less important so that tides began to dominate the evolution. For many close-in planets, Figure 8 shows that initial $a$ values were significantly higher than the currently observed values. These initial values of $a$ likely represent their locations at the termination of gas disk migration in each early planetary system.

Figure 9 indicates where that gas disk migration may have halted for different choices of $Q$ values. In that figure, we plot, as a function of $Q_*$ and $Q_p$, the inner edge of the initial $a$ distribution, defined here as second smallest $a_{initial}$ for our group of modeled planets as a function of $Q_*$ and $Q_p$. (We use the *second* smallest $a_{initial}$ because, in looking for a trend for $a_{initial}$, we might expect at least one outlier to skew the trend.) The inner edge of $a_{initial}$ values may represent the distance at which gas disk migration halted for the planets. For our best-fit $Q$'s, the edge is at $a_{initial}$ = 0.037 AU (GJ 436 b), about twice as far out as the smallest current $a$ values (Figure 1).

## 5. Discussion

This investigation supports the hypothesis that tidal interactions between a star and a planet are responsible for the relatively small $e$ values of close-in planets (as proposed by Rasio *et al.* 1996), although our calculations incorporate important corrections to previous studies. We have used a more complete set of tidal evolution equations, including tides raised on both planet and star, considering the strong coupling between eccentricity and semi-major axis evolution, and avoiding inappropriate assumptions such as conservation of angular momentum. Going back in time, the eccentricity distribution is restored to match that of farther-out planets.

Even with the relatively small eccentricities of the close-in planets explained, there remains the unresolved question of the origin of the initial eccentricity distribution of both close-in and farther-out planets. Rasio & Ford (1996) and Weidenschilling & Marzari (1996) originally proposed that planet-planet scattering early in the life of an extra-solar system could naturally lead to the large eccentricities observed. Subsequent work seems to support this initial hypothesis (Rasio

& Ford 1996; Ford, Havlikova, & Rasio 2001; Marzari & Weidenschilling 2002; Ford & Rasio 2005). An alternative model to explain the eccentricity distribution is that the eccentricities were set during the formation of planets in a gaseous disk. In this scenario planets excite spiral density waves in the disk, and the subsequent gravitational interactions between the planet and these waves can increase eccentricity. However, models of disk migration appear unable to pump up eccentricities to large (> 0.3) values (*i.e.*, Boss 2000; D'Angelo *et al.* 2006). Since our work has shown even close-in planets formed with much larger *e*, the most plausible mechanism for producing the initial *e* values is the planet-planet scattering model, although some modification to the original views of that model are needed (see, *e.g.*, Barnes & Greenberg 2007). Thus the origin of the large and widely distributed eccentricities among extra-solar planets remain uncertain.

Significant reductions in semi-major axes have accompanied the changes in eccentricity, with important implications. First, models of protoplanetary migration in the primordial gas disk need not carry "hot Jupiters" in as far as their current positions. Lin *et al.* (1996) proposed that migration in the gas disk halted near the inner edge of the disk, a boundary determined by clearing due to the host star's magnetosphere. Our results show that the inner edge was probably farther out than indicated by the current semi-major axes of the planets, which were only reached during tidal migration long after the nebula had dissipated. In order to evaluate where migration due to the gas disk halted, (and thus where the inner edge of the nebula was) models should account for the subsequent tidal evolution.

The tidal changes in orbital semi-major axes also have implications for observations of planetary transits, such as surveys of young open galactic clusters (Bramich *et al.* 2005; Burke *et al.* 2006). The probability to observe a planetary transit increases for smaller semi-major axes, but decreases as orbits become more circular (Borucki et al.1984; Charbonneau et al. 2007; Barnes 2007). Tidal evolution means that the probability of an observable transit depends on a star's age, but the exact relation depends on the particular evolutionary path through ($a$,$e$) space. As our understanding of the statistics of tidal evolution paths improves, the observed frequency of transits in the field and in open clusters may eventually help to constrain planetary formation scenarios, distinguishing for example between the relative roles of embedded migration and of gravitational scattering, which set up the initial conditions for tidal evolution. However, transit statistics may not yet be refined enough to be sensitive to detect this effect (Pepper & Gaudi 2005).

The inward evolution of semi-major axes of close-in planets also has implications for their thermal history because the solar energy received by the planets over their lifetimes may have been less than assumed. For example, Hubbard *et al.* (2007) studied the effects of evaporative mass loss from the close-in extra-solar planets. They calculated that the smaller planets should lose mass more quickly than larger ones. However, the mass-frequency distribution is fairly independent of $a$. They concluded that mass loss by evaporation has been negligible, and that any migration in $a$ has been independent of mass. However, our results indicate that during tidal evolution, more massive planets have probably experienced significantly more migration than less massive ones, and interpretation of mass distributions should be revisited in light of tidal evolution.

The tidally driven changes in semi-major axes also may affect modeling of interior processes in the planets and the implications for their radii. Burrows *et al.* (2007) modeled the effects of radiative cooling and stellar insolation in an attempt to understand the observed radii of transiting planets and their implications for internal structure and atmospheric opacity. Since tides have probably induced migration for many close-in planets from initially more distant orbits, Burrows *et al.* overestimated the total insolation experienced by these planets. Revision of these models taking into account a variable insolation history may give somewhat smaller planetary radii, which may exacerbate disagreement between the theory and radii inferred from transit observations.

On the other hand, tidal heating that accompanied the orbital evolution would have added to the heat budget considered by Burrows *et al.* and might have significant effects on planetary radii, as suggested by Bodenheimer *et al.* (2003). Especially in the past, when $e$ was much larger, tidal heating may have contributed significantly to planetary thermal budgets. This source of heat may be important for reconsidering evaporative mass loss, as well as interior structure. Tidal heating rates can be derived directly from the results presented here (Jackson *et al.* 2007).

## 6. Conclusions

The results presented here confirm that tidal evolution is likely responsible for the unusually low orbital eccentricities of close-in extra-solar planets. The initial distribution of $e$ for these planets (at the time that tidal evolution became dominant, after the gas disk had dissipated) agrees well with the $e$ distribution for farther-out planets. The best agreement is with $Q_* = 10^{5.5}$ and $Q_p = 10^{6.5}$. These $Q$ values are consistent with estimates from orbital constraints in the Solar System, as well as from models of the physics of gaseous bodies.

Our numerical solutions of the tidal evolution equations demonstrate the importance of retaining all terms representing the tides on both the planet and the star. It is inaccurate, when considering changes in eccentricities, to ignore tides raised on the star. Similarly, it is inaccurate, when considering changes in semi-major axis, to ignore tides raised on the planet. Incorporating these effects significantly improves the accuracy of our results relative to previous studies. We have also shown the significance of the coupling between the evolution of $a$ and $e$. Previous conclusions reached by considering the “timescale” for changes in $e$ have depended on the assumption that the simultaneous changes in $a$ can be ignored. Using that approximation, the time dependence of $e$ would be exponential, with a corresponding timescale. However, the simultaneous changes in $a$ cause the time variation of $e$ to be more complex, with $de/dt$ not even necessarily changing monotonically over the age of a planet. Similarly, “tidal migration” has generally been considered to be a process of change in semi-major axis. However, such change is coupled intimately with the changes in $e$, so tidal migration should always refer to both aspects of the change in an orbit. In summary, the results of studies that incorporated approximate solutions to Equations (1) and (2) should be applied with caution.

Our full treatment of tidal evolution has revealed that close-in planets have current orbital semi-major axes typically half as large as they were at the time that the planets had formed and the nebula cleared. Even the closest-in planets probably started with $a_{initial} > 0.04$ AU.

We have neglected interactions between the tidally evolving planet and other planets in its system. Ten of the 36 of the planets in our study are known to be members of a multi-planet system. Models that couple tidal evolution to planet-planet interactions have been developed (Mardling & Lin 2002; Adams & Laughlin 2006), but they require specific assumptions about the tidal response of bodies. The modeling of such complex phenomenon was beyond the scope of this paper, but future work should include these effects.

Since our equations depend sensitively on many uncertain parameters (such as system age, $a_{current}$, and planetary radii), the exact model solution we present here is not exact for any specific planet. However, assuming that errors in our choices for significant parameters are evenly spread about their correct values, our calculations probably provide a reasonable representation of the population as a whole. Similarly, our assumption that all planets have the same $Q_p$ as one another, and that all stars have the same $Q_*$ as one another, surely introduces errors for individual planets and stars, but probably provides a reasonable first estimate of the evolution of the population as a whole.

In the future, our approach can be revised and updated, and the results can be refined. As tidal dissipation mechanisms come to be better understood, particularly their dependence on frequency, we can improve our calculations, and use forms of the governing equations that are more reliable at large $e$. As more extra-solar planets are discovered, the quality of our statistical analysis will improve. Continued radial-velocity observations will result in better orbital solutions, while transit observations will provide us with more accurate planetary masses and radii. Also, improvements in stellar age estimates will help us to better understand the duration of tidal effects. This study lays the groundwork for future investigations into the nature of extra-solar tidal evolution.

Acknowledgments:  We thank Adam Burrows, Doug Hamilton, Maki Hattori, William Hubbard, Steve Lubow, Gordon Ogilvie, Sean Raymond, Genya Takeda and Ke Zhang for useful discussions and suggestions. This research was supported by the NASA PG&G program.

TABLE 1 – Extra-Solar Planet and Host Star Data

| Name | M* (solar masses) | R* (solar radii) | $M_p$ (Jupiter masses) | $R_p$ (Jupiter radii) | $a_{current}$ (AU) | $e_{current}$ | Nominal Age (Gyr) | Minimum Age (Gyr) | Maximum Age (Gyr) |
|---|---|---|---|---|---|---|---|---|---|
| 51 Peg b | 1.09[2] | 1.15[2] | 0.472[2] | 1.20 | 0.05[2] | 0.01[2] | 6.76[18] | 5.28[18] | 8.40[18] |
| 55 Cnc b | 0.92[2] | 0.93[2] | 0.833[14] | 1.20 | 0.11[14] | 0.01[14] | … | 7.24[18] | … |
| 55 Cnc e | 0.92[2] | 0.93[2] | 0.038[14] | 0.34 | 0.04[14] | 0.09[14] | … | 7.24[18] | … |
| BD 10-3166 b | 0.92[2] | 0.84[2] | 0.458[2] | 1.20 | 0.05[2] | 0.02[2] | … | … | 1.84[18] |
| GJ 436 b | 0.44[12] | 0.46[6] | 0.077[2] | 0.37[4] | 0.03[16] | 0.15[16] | … | 7.41[17] | 11.05[17] |
| GJ 876 c | 0.32[18] | 0.39[7] | 0.619[16] | 1.20 | 0.13[16] | 0.22[16] | … | 6.52[17] | 9.90[17] |
| HAT-P-2 b | 1.928[1] | 1.474[1] | 9.04[1] | 0.982[1] | 0.07[1] | 0.52[1] | 2.60[1] | 1.20[1] | 3.40[1] |
| HD 102117 b | 1.11[2] | 1.26[2] | 0.170[2] | 0.55 | 0.15[2] | 0.09[2] | 9.40[18] | 8.04[18] | 10.60[18] |
| HD 108147 b | 1.19[2] | 1.25[2] | 0.261[2] | 0.64 | 0.10[2] | 0.53[2] | 3.20[18] | 2.32[18] | 3.92[18] |
| HD 118203 b | 1.23[3] | 2.13[3] | 2.140[3] | 1.20 | 0.07[3] | 0.31[3] | 4.60[3] | 3.80[3] | 5.40[3] |
| HD 130322 b | 0.88[2] | 0.85[2] | 1.090[2] | 1.20 | 0.09[2] | 0.03[2] | … | 10.80[18] | … |
| HD 13445 b | 0.77[2] | 0.80[2] | 3.910[2] | 1.20 | 0.11[2] | 0.04[2] | … | 8.48[18] | … |
| HD 149143 b | 1.10[3] | 1.55[3] | 1.330[2] | 1.20 | 0.05[3] | 0.08[3] | 7.60[3] | 6.40[3] | 8.80[3] |
| HD 162020 b | 0.78[19] | 0.74[19] | 15.000[2] | 1.20 | 0.08[19] | 0.28[19] | … | … | 0.76[18] |
| HD 168746 b | 0.93[2] | 1.04[2] | 0.248[2] | 0.63 | 0.07[2] | 0.11[2] | 12.40[18] | 10.28[18] | … |
| HD 179949 b | 1.21[2] | 1.22[2] | 0.916[2] | 1.20 | 0.04[2] | 0.02[2] | 2.56[18] | 0.92[18] | 3.68[18] |
| HD 185269 b | 1.28[8] | 1.88[8] | 0.909[2] | 1.20 | 0.08[8] | 0.28[8] | 4.00[15] | 3.00[15] | … |
| HD 187123 b | 1.08[2] | 1.17[2] | 0.532[2] | 1.20 | 0.04[2] | 0.04[2] | 7.40[18] | 6.24[18] | 8.64[18] |
| HD 192263 b | 0.81[2] | 0.77[2] | 0.641[2] | 1.20 | 0.15[2] | 0.06[2] | 2.56[18] | … | 13.36[18] |
| HD 195019 b | 1.07[2] | 1.38[2] | 3.700[2] | 1.20 | 0.14[2] | 0.01[2] | 2.00[18] | … | 7.04[18] |
| HD 209458 b | 1.14[2] | 1.14[2] | 0.690[2] | 1.32[9] | 0.05[10] | 0.01[10] | 2.44[18] | 0.80[18] | 3.76[18] |
| HD 217107 b | 1.10[2] | 1.08[2] | 1.410[20] | 1.20 | 0.07[20] | 0.13[20] | 5.84[18] | 3.40[18] | 7.76[18] |
| HD 38529 b | 1.47[2] | 2.50[2] | 0.852[2] | 1.20 | 0.13[2] | 0.25[2] | 3.28[18] | 3.04[18] | 3.64[18] |
| HD 41004B b | 0.40[22] | 0.49[7] | 18.400[22] | 1.20 | 0.02[22] | 0.08[22] | … | 1.48[17] | 1.64[17] |
| HD 46375 b | 0.92[2] | 0.94[2] | 0.226[2] | 0.61 | 0.04[2] | 0.06[2] | … | 11.88[18] | … |
| HD 49674 b | 1.06[2] | 0.95[2] | 0.115[2] | 0.49 | 0.06[2] | 0.29[2] | … | .. | 3.56[18] |
| HD 6434 b | 0.79[13] | 0.74 | 0.397[13] | 1.20 | 0.14[13] | 0.17[13] | 13.30[17] | 7.00[17] | … |
| HD 68988 b | 1.18[2] | 1.14[2] | 1.850[21] | 1.20 | 0.07[21] | 0.12[21] | 3.40[18] | 1.40[18] | 4.44[18] |
| HD 69830 b | 0.87[2] | 0.90[2] | 0.032[11] | 0.32 | 0.08[11] | 0.10[11] | … | 12.04[18] | … |
| HD 75289 b | 1.21[2] | 1.28[2] | 0.467[2] | 1.20 | 0.05[2] | 0.03[2] | 3.28[18] | 2.60[18] | 3.88[18] |
| HD 76700 b | 1.13[2] | 1.34[2] | 0.233[2] | 0.62 | 0.05[2] | 0.10[2] | 9.84[18] | 8.80[18] | 12.76[18] |
| HD 83443 b | 1.00[2] | 1.02[2] | 0.398[2] | 1.20 | 0.04[2] | 0.01[2] | … | 11.68[18] | … |
| HD 88133 b | 1.20[5] | 1.93[5] | 0.299[2] | 0.67 | 0.05[2] | 0.13[2] | … | 6.27[18] | 9.56[18] |
| HD 99492 b | 0.86[2] | 0.76[2] | 0.109[2] | 0.48 | 0.12[2] | 0.25[2] | … | … | 1.80[18] |
| τ Boo b | 1.35[2] | 1.43[2] | 4.130[2] | 1.20 | 0.05[2] | 0.02[2] | 1.64[18] | 1.12[18] | 2.08[18] |
| υ And b | 1.32[2] | 1.42[2] | 0.687[2] | 1.20 | 0.06[2] | 0.02[2] | 3.12[18] | 2.88[18] | 3.32[18] |

Note: 1 solar mass = $1.9891\times10^{30}$ kg, 1 solar radius = 695500 km, 1 Jupiter mass = $1.8986\times10^{27}$ kg, and 1 Jupiter radius = 71492 km. $R_p$ values were computed as discussed in the text.

References: (1) Bakos *et al.* (2007); (2) Butler *et al.* (2006), (3) Da Silva *et al.* (2006); (4) Deming *et al.* (2007); (5) Fischer & Valenti (2005); (6) Gillon *et al.* (2007); (7) Gorda & Svechnikov (1996); (8) Johnson *et al.* (2006); (9) Knutson *et al.* (2007); (10) Laughlin *et al.* (2005c); (11) Lovis *et al.* (2006); (12) McArthur *et al.* (2004); (13) Maness *et al.* (2007); (14) Mayor *et al.* (2004); (15) Moutou *et al.* (2006); (16) Rivera *et al.* (2005); (17) Saffe *et al.* (2006); (18) Takeda *et al.* (2007); (19) Udry *et al.* (2002); (20) Valenti & Fischer (2005); (21) Vogt *et al.* (2005); (22) Wright *et al.* (2006); (23) Zucker *et al.* (2004).

Figure Captions

Figure 1: Distribution of extra-solar planetary eccentricities *e* and semi-major axes *a*. Average *e* is 0.09 for $a \leq 0.2$ AU and about 0.3 for $a > 0.2$ AU. This disparity is widely attributed to tidal circularization for the close-in orbits through interactions between the planet and the host star. The data for Figure 1 are taken from Butler et al 2006, with supplemental data from Udry *et al.* (2002), MacArthur *et al.* (2004), Mayor *et al.* 2004, Zucker *et al.* (2004), Laughlin *et al.* (2005b), Rivera *et al.* (2005), Vogt *et al.* (2005), Da Silva *et al.* (2006), Johnson *et al.* (2006), Lovis *et al.* (2006), Wright *et al.* (2006), and Bakos *et al.* (2007).

Figure 2: Circularization timescale $\tau_{\rm circ}$ as a function of planetary mass $M_p$ for HD 217107 b, estimated to have an age ~ 6 Gyr. The line labeled "$\tau_{\rm circ}$ w/o stellar tide" is the eccentricity circularization timescale. In this case, $M_p$ must be > 4 $M_J$ for the large *e* to be preserved ($\tau_{\rm circ} > \tau_{\rm age}$). The line labeled "$\tau_{\rm circ}$ w/ stellar tide" includes the effect of the tide raised on the star (but still assuming constant *a*), which decreases $\tau_{\rm circ}$ as $M_p$ grows, such that there is no value of $M_p$ for which $\tau_{\rm circ} > \tau_{\rm age}$.

Figure 3: Tidal evolution of *e* and *a* for τ Boo b backwards in time from their current values ($e_{current}$ = 0.023, $a_{current}$ = 0.0595 AU). Here $Q_p$ is fixed at $10^5$, and results are shown for various $Q_*$ values. Solid lines are numerical integration of Equation (1) and (2). The dashed lines show the exponential solutions to Equation (1) with *a* assumed constant, a common but inappropriate assumption. In the left panel, note that the lines labeled $Q_* = 10^4$ and $Q_* = 10^6$ are, in fact, correctly labeled. See text for details.

Figure 4: A contour plot of the dependence of $e_{initial}$ for τ Boo b on $Q_*$ and $Q_p$. The current value of $e$ is 0.023. The "heel" in the contour lines for $10^4 < Q_* < 10^6$ illustrates the counter-intuitive relationship of the change in $e$ with $Q_*$: Enhancing stellar tidal dissipation (*i.e.* smaller $Q_*$) can result in a decreasing amount of change of $e$. This behavior is not unique to τ Boo b.

Figure 5: Distributions of orbital elements for four examples of pairs of $Q_p$ and $Q_*$ values out of the 289 we tested. Squares (filled and open) represent the currently observed orbital elements, with the open squares (with $a < 0.2$ AU) being candidates for significant tidal evolution. Triangles show the initial orbital elements ($e_{initial}$ and $a_{initial}$) determined by integrating the equations of tidal evolution backward in time to the formation of the planet.

Figure 6: A contour plot of K-S test score as a function of $Q_*$ and $Q_p$ for comparison of the computed initial $e$ distribution (after integrating tidal evolution back in time) of close-in planets ($a < 0.2$ AU), with the current $e$ distribution for planets with $a < 0.2$ AU. Contours are spaced in 0.1 intervals, and local maxima are indicated with X's. Moving from top to bottom, local maxima correspond to K-S scores of 0.77, 0.89, and 0.76, respectively.

Figure 7: Tidal evolution of $e$ and $a$ for the sample of known close-in extra-solar planets using our best-fit values of $Q_* = 10^{5.5}$ and $Q_p = 10^{6.5}$. Solid curves represent the trajectories of orbital evolution from current orbits (lower left end of each curve) backward in time (toward the upper right). On the trajectories, tick marks are spaced every 500 Myr to indicate the rate of tidal evolution. Tidal integrations were performed for 15 Gyr for all planets, but the filled circles

indicate the initial values of orbital elements at the beginning of each planet's life. Due to space restrictions, most planets are not labeled, however they can be identified by the ($a$, $e$) values at the lower left end of each trajectory, using Table 1. For example, 51 Peg b starts at ($a$, $e$) = (0.0527 AU, 0.013).

Figure 8: Change in $a$ for each close-in planet due to tidal evolution from the initial value to the current value. Results are shown for the same four pairs of $Q$ values as in Figure 5. The distance above the diagonal line ($a_{current} = a_{initial}$) represents the distance a planet has migrated inward (in $a$) since the nebula dissipated and tides became the dominant effect.

Figure 9: The inner edge of the $a$ distribution for inferred initial conditions ($a_{initial}$ in AU), as a function of the assumed values of $Q_*$ and $Q_p$. The "inner edge" is the second smallest value of $a_{initial}$. (The smallest value is assumed to be a less significant outlier.) The current edge by this definition is $a_{current}$ = 0.0278 AU. For our best-fit $Q$ values (shown by the X), the inner edge was about twice as far from the star as the current position.

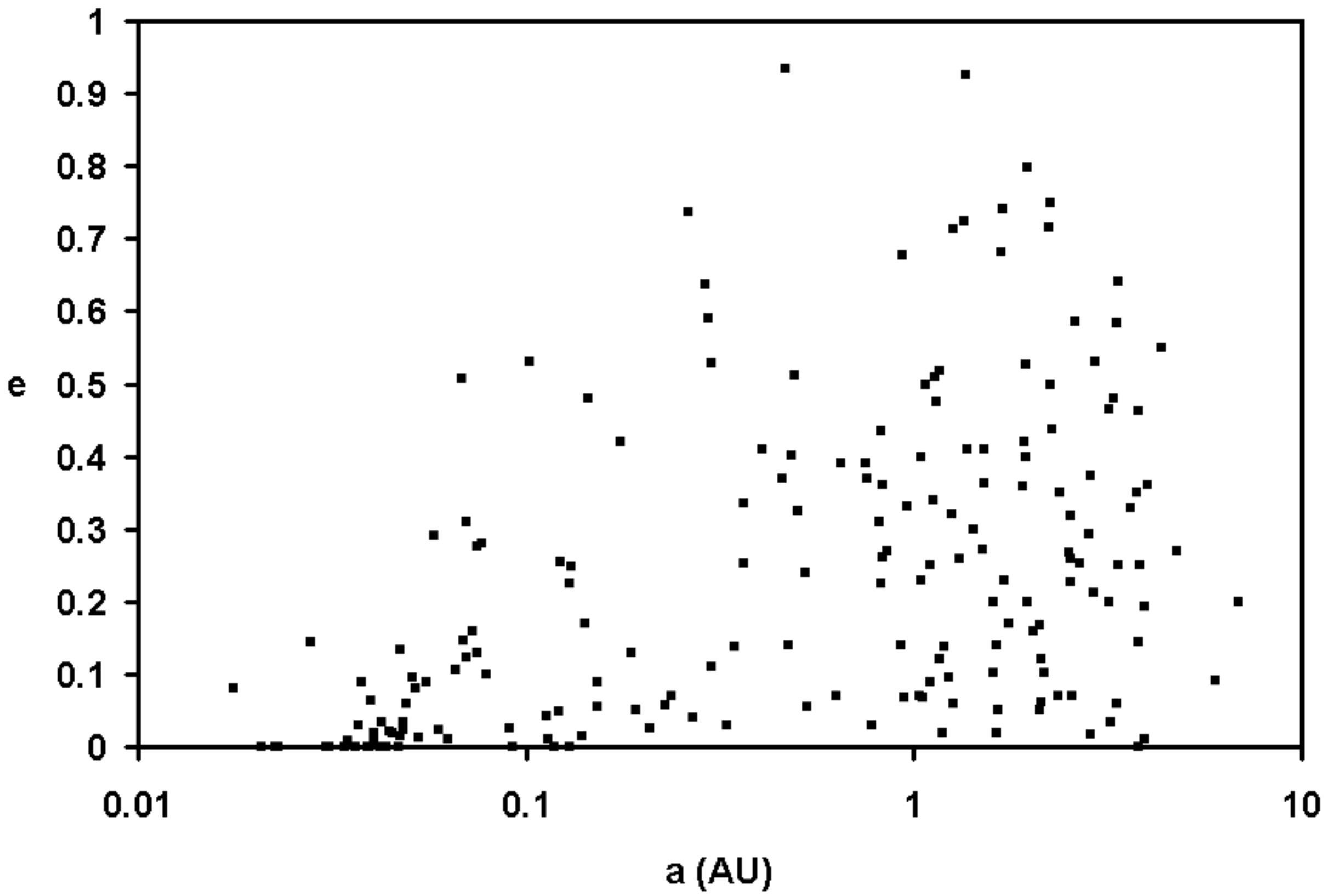

Figure 1

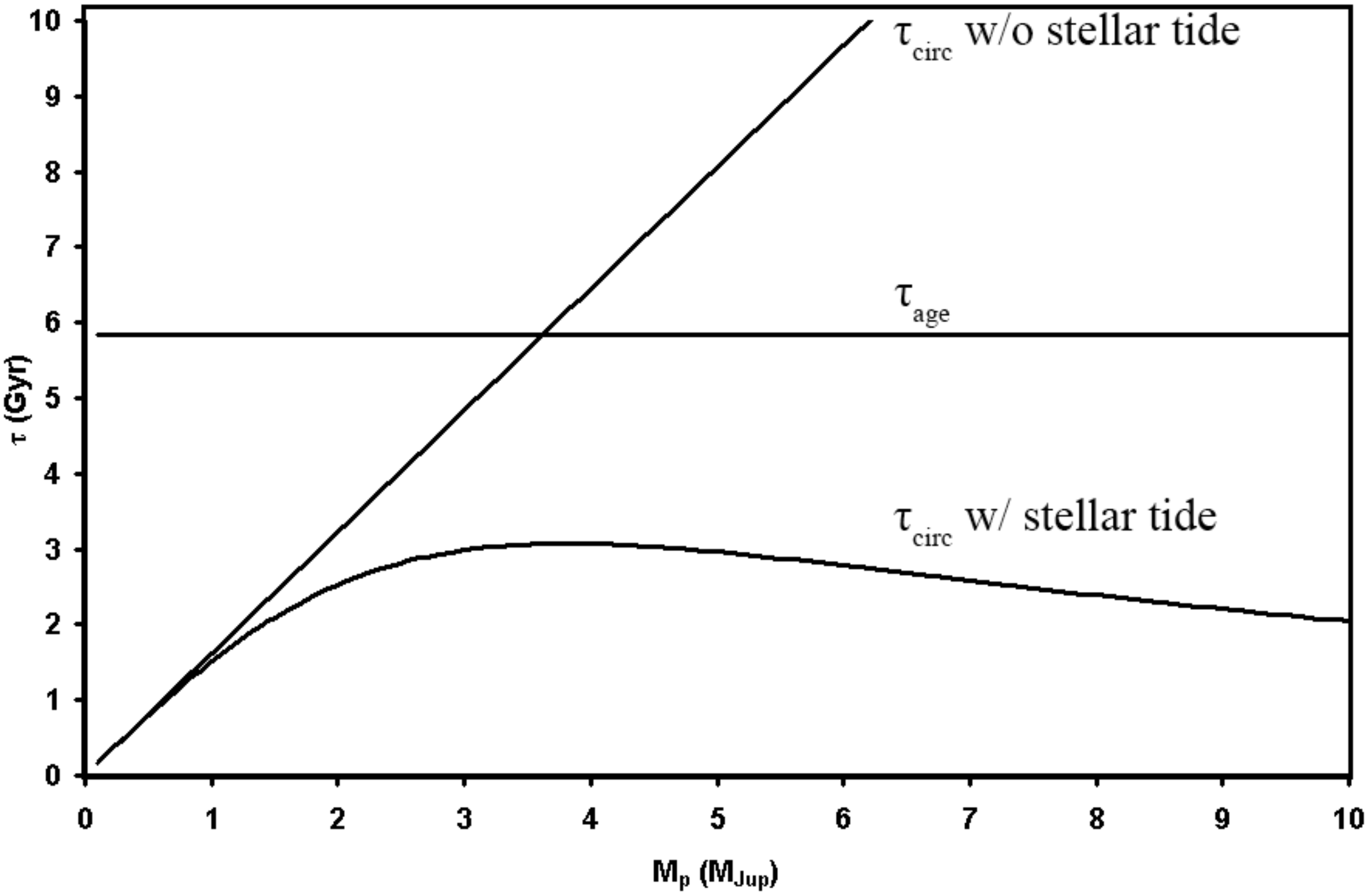

Figure 2

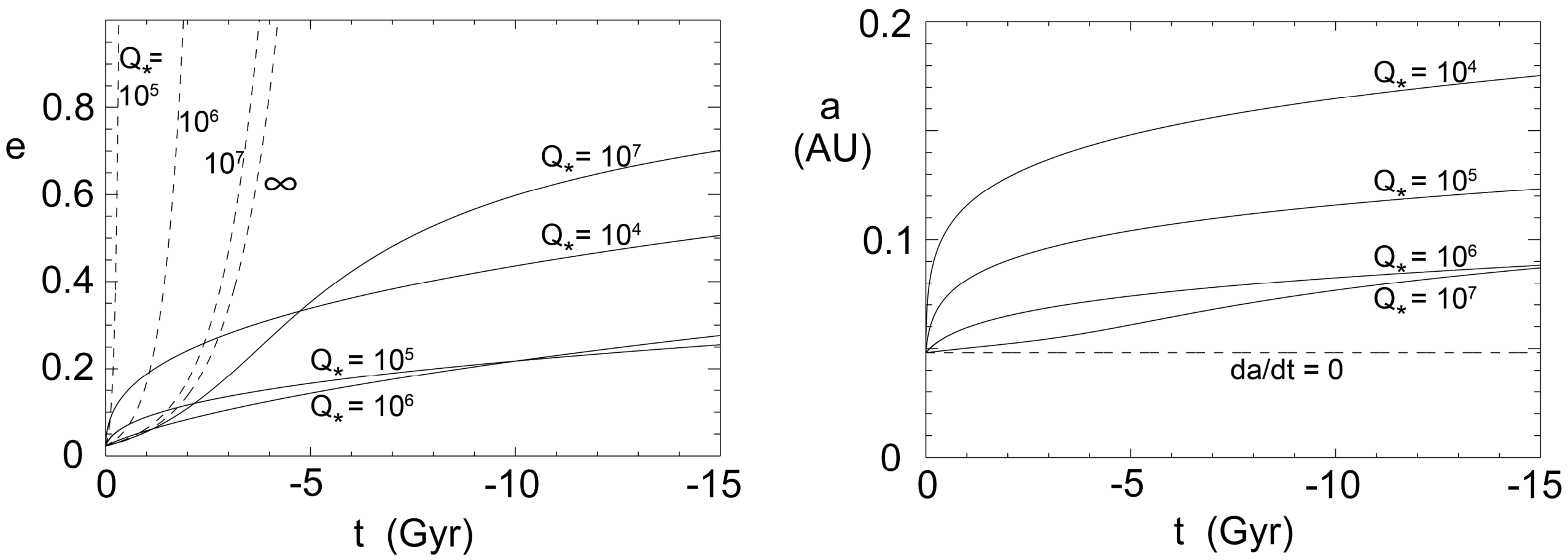

Figure 3

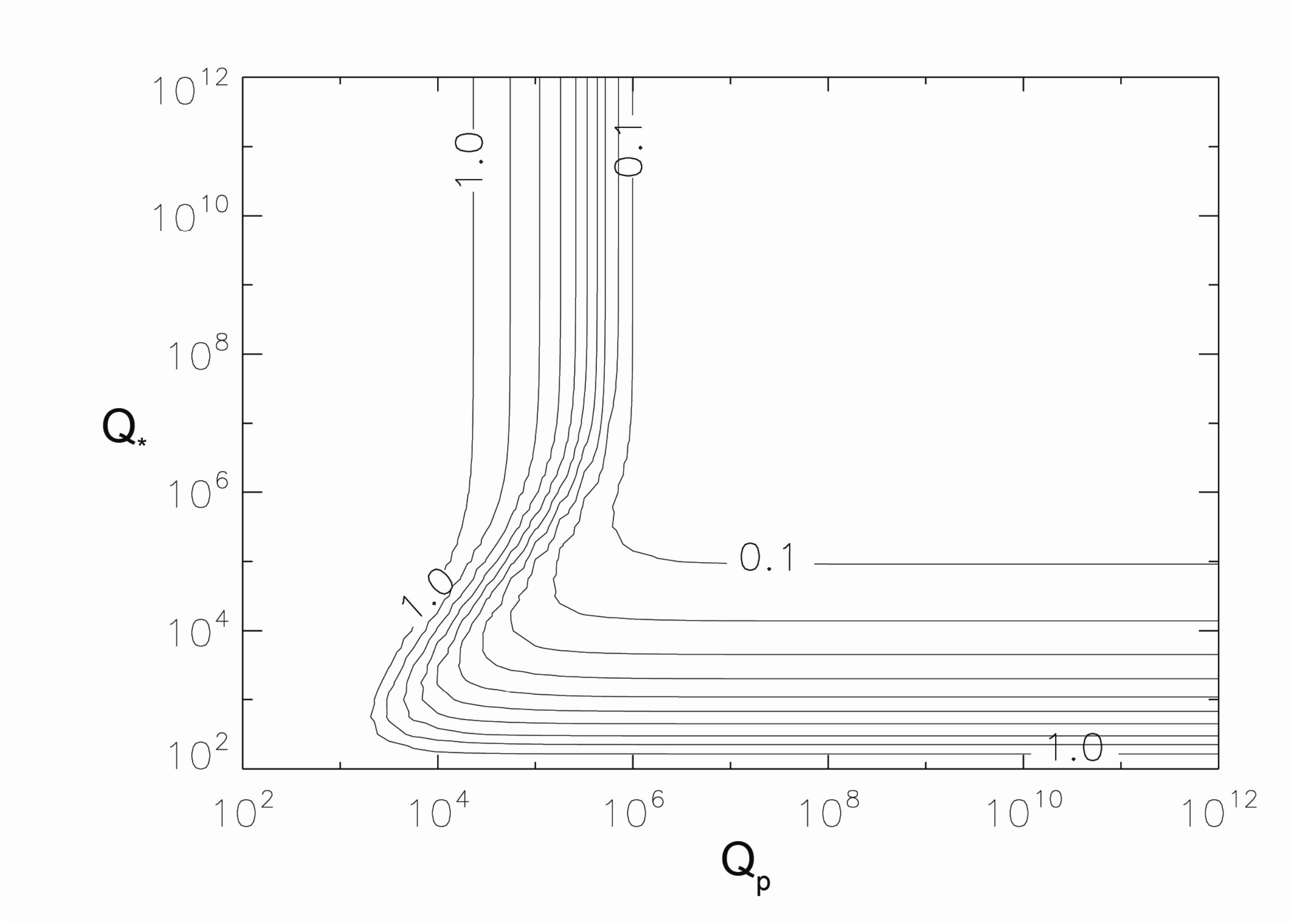

Figure 4

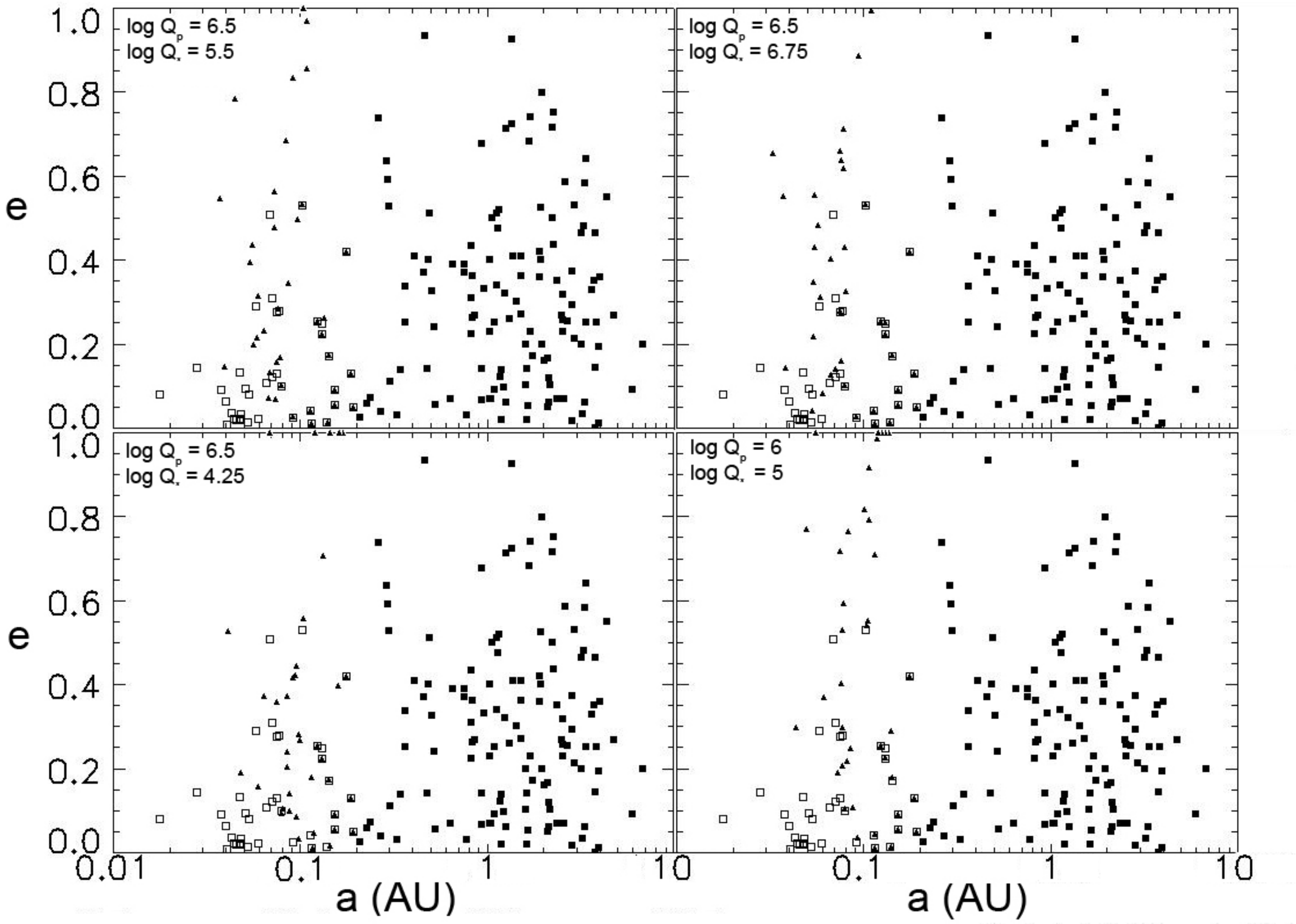
log $Q_p$ = 6.5
log $Q_*$ = 5.5
log $Q_p$ = 6.5
log $Q_*$ = 6.75
log $Q_p$ = 6.5
log $Q_*$ = 4.25
log $Q_p$ = 6
log $Q_*$ = 5
e
1.0
0.8
0.6
0.4
0.2
0.0
0.01
0.1
1
10
a (AU)

Figure 5

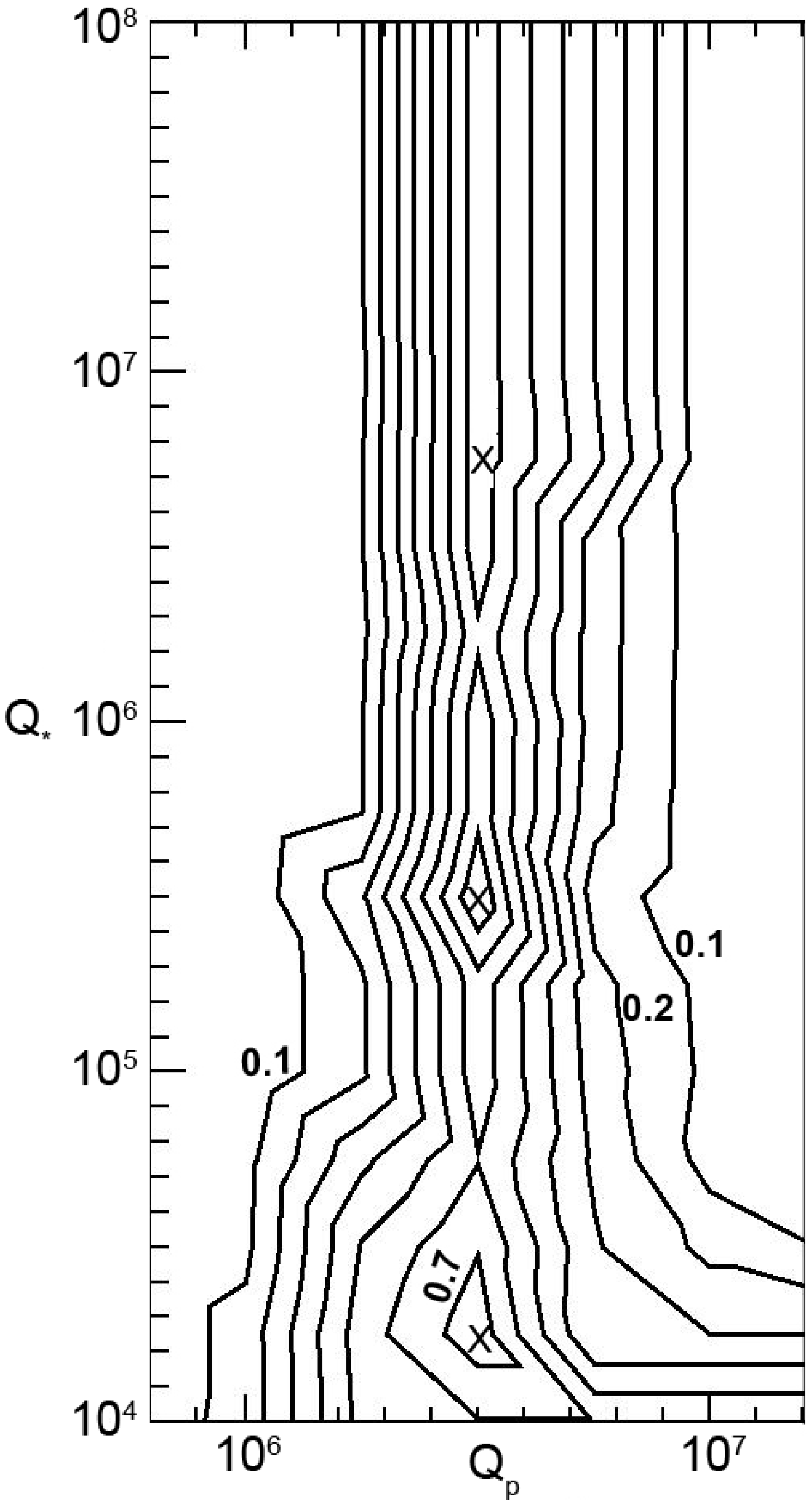

Figure 6

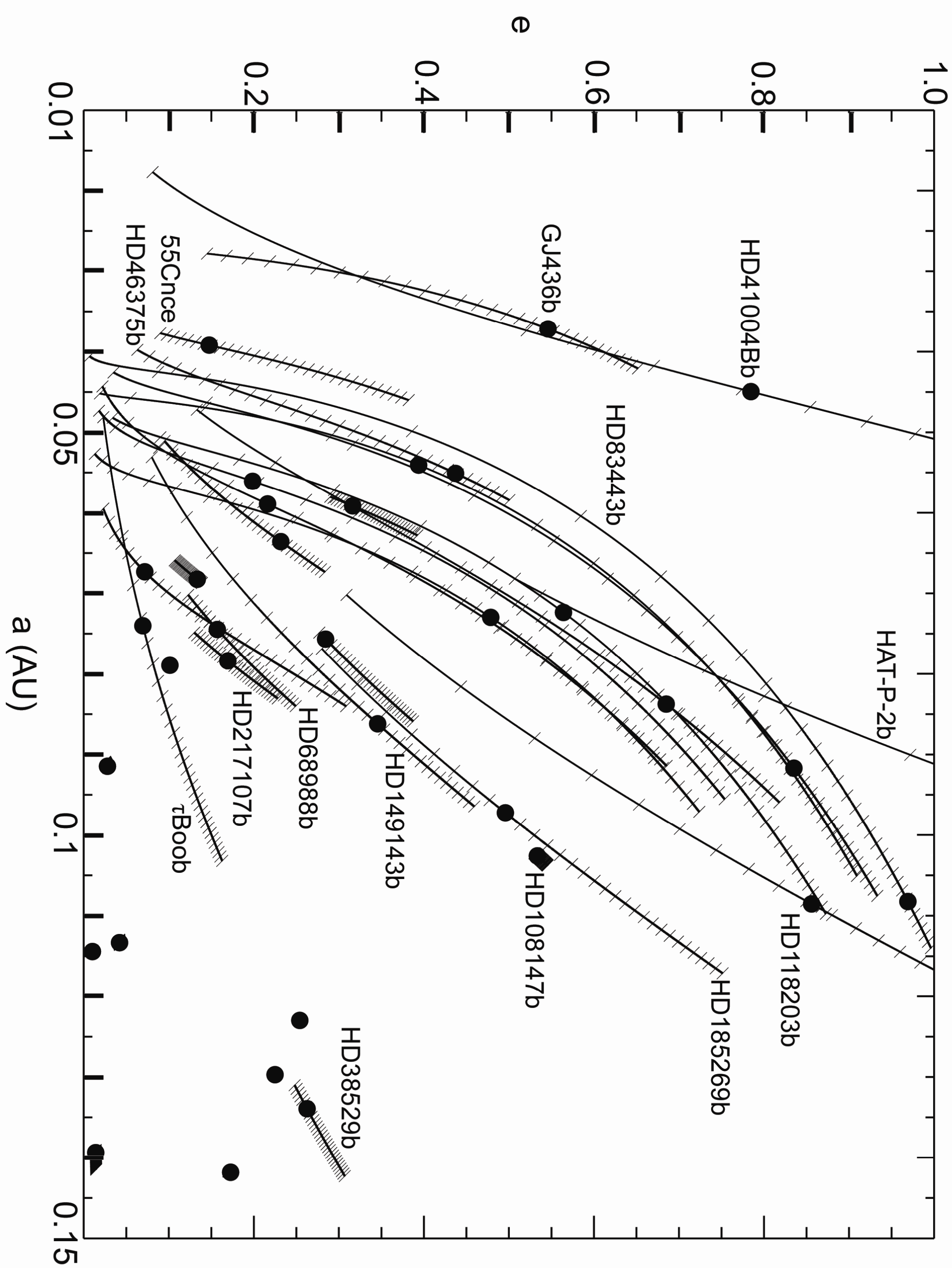

Figure 7

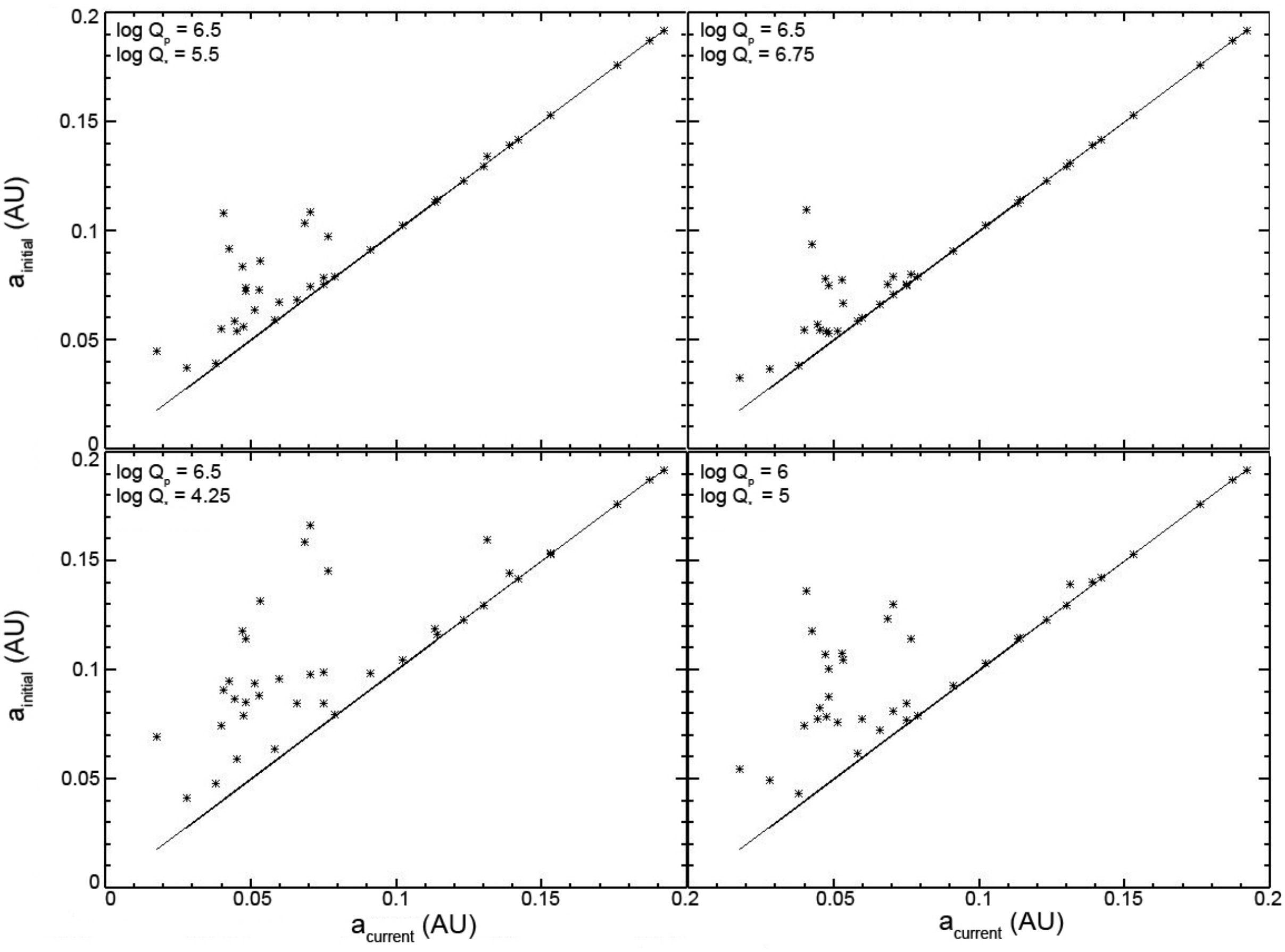
log $Q_p$ = 6.5
log $Q_*$ = 5.5
log $Q_p$ = 6.5
log $Q_*$ = 6.75
log $Q_p$ = 6.5
log $Q_*$ = 4.25
log $Q_p$ = 6
log $Q_*$ = 5
$a_{initial}$ (AU)
$a_{current}$ (AU)
0
0.05
0.1
0.15
0.2

Figure 8

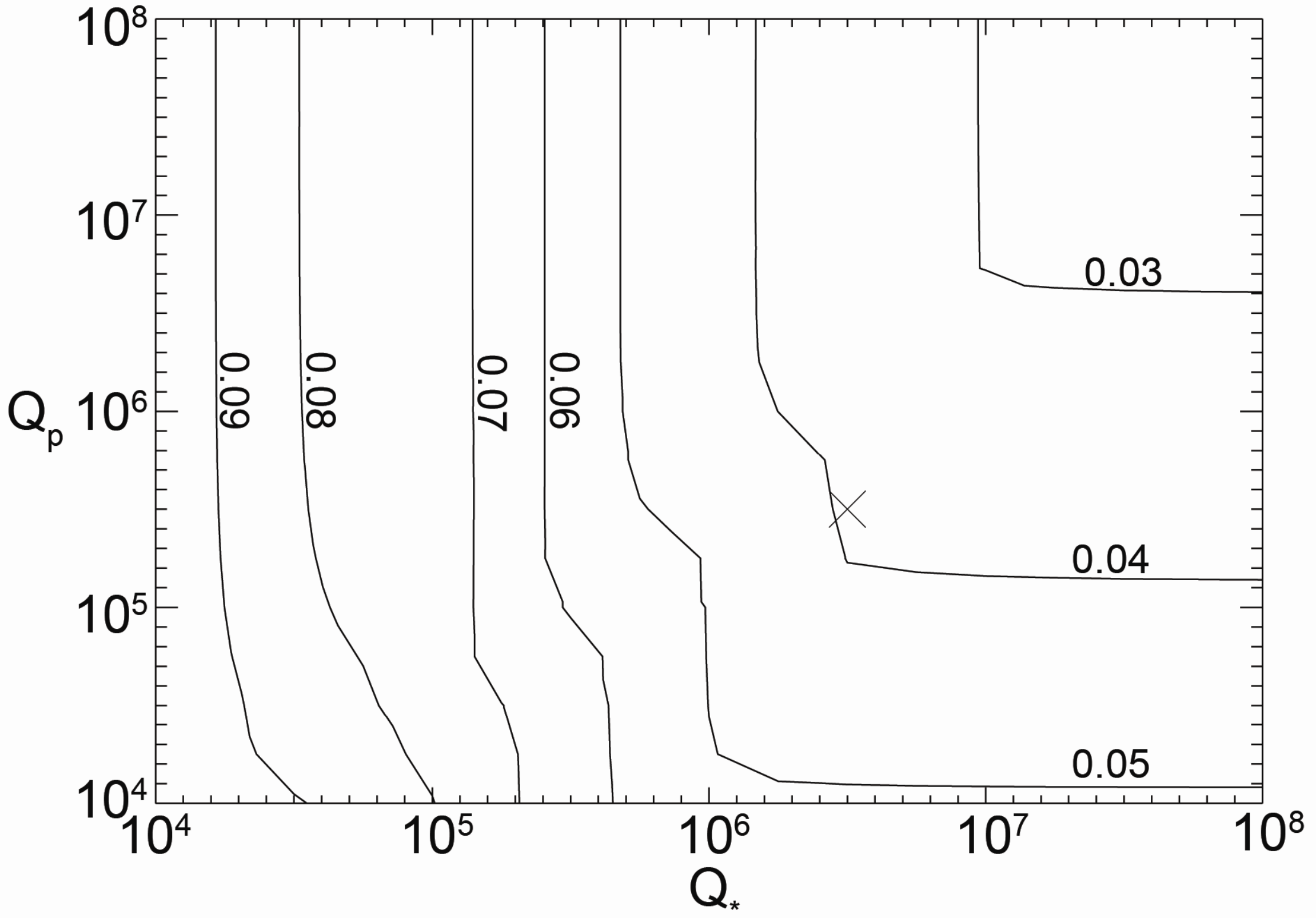

Figure 9